\documentclass[journal,twcolumn]{IEEEtran}
\usepackage{cite}
\usepackage[cmex10]{amsmath}
\usepackage{amsthm}
\usepackage{amssymb}
\usepackage{balance}
\usepackage[final]{graphicx}
\usepackage{color}
\usepackage{epsfig}
\usepackage{epstopdf}
\usepackage{subcaption}
\usepackage{multirow}
\usepackage{tikz}
\tikzset{>=latex}
\usetikzlibrary{positioning}
\usepackage[ruled,vlined,linesnumbered]{algorithm2e}

\newcommand{\bs}[1]{\boldsymbol{#1}}
\newcommand{\mc}[1]{\mathcal{#1}}

\newcommand{\mb}[1]{\mathbf{#1}}

\definecolor{LayerColor}{RGB}{230, 230, 230}
\definecolor{InOutColor}{RGB}{240, 243, 255}
\definecolor{cellColor}{RGB}{230, 230, 230}

\definecolor{kugray5}{RGB}{224,224,224}

\usepackage[normalem]{ulem}
\newcommand\rsout{\bgroup\markoverwith
	{\textcolor{red}{\rule[0.5ex]{2pt}{0.8pt}}}\ULon}

\usepackage{array,ragged2e}
\newcolumntype{P}[1]{>{\RaggedRight\arraybackslash}p{#1}}



\usepackage{etoolbox}
\let\mybibitem\bibitem
\renewcommand{\bibitem}[1]{%
	\ifstrequal{#1}{nature}
	{\color{blue}\mybibitem{#1}}
	{\color{black}\mybibitem{#1}}%
}

\graphicspath{ {Figures/} }

\newcommand{\norm}[1]{\left\lVert#1\right\rVert} % ||.||
\newcommand{\setA}{\mathcal{A}}

\begin{document}

\title{Leveraging Deep Neural Networks for Massive MIMO Data Detection}
%\author{\copyright SDSU}
\author{Ly~V.~Nguyen$^*$, Nhan~T.~Nguyen$^*$, Nghi~H.~Tran, Markku Juntti, A.~Lee~Swindlehurst, and Duy~H.~N.~Nguyen
	\thanks{Ly V. Nguyen is with the Computational Science Research Center, San Diego State University, San Diego, CA, USA 92182 (e-mail: vnguyen6@sdsu.edu).}
    \thanks{Nhan Thanh Nguyen and Markku Juntti are with Centre for Wireless Communications, University of Oulu, P.O.Box 4500, FI-90014, Finland, (email: nhan.nguyen@oulu.fi, markku.juntti@oulu.fi).}
	\thanks{Nghi H. Tran is with the Department of Electrical and Computer Engineering, University of Akron, OH, USA 44325 USA (e-mail: nghi.tran@uakron.edu)}
	\thanks{A. Lee Swindlehurst is with the Center for Pervasive Communications and Computing, Henry Samueli School of Engineering, University of California, Irvine, CA, USA 92697 (e-mail: swindle@uci.edu).}
	\thanks{Duy H. N. Nguyen is with the Department of Electrical and Computer Engineering, San Diego State University, San Diego, CA, USA 92182 (e-mail: duy.nguyen@sdsu.edu).}
	\thanks{$^*$These authors contribute equally to this work.}}
%	\thanks{\emph{Corresponding author:} Duy H. N. Nguyen.}}

\maketitle

\begin{abstract}
Massive multiple-input multiple-output (MIMO) is a key technology for emerging next-generation wireless systems. Utilizing large antenna arrays at base-stations, massive MIMO enables substantial spatial multiplexing gains by simultaneously serving a large number of users. However, the complexity in  massive MIMO signal processing (e.g., data detection) increases rapidly with the number of users, making conventional hand-engineered algorithms less computationally efficient. Low-complexity massive MIMO detection algorithms, especially those inspired or aided by deep learning, have emerged as a promising solution. While there exist many MIMO detection algorithms, the aim of this magazine paper is to provide insight into how to leverage deep neural networks (DNN) for massive MIMO detection. We review recent developments in DNN-based MIMO detection that incorporate the domain knowledge of established MIMO detection algorithms with the learning capability of DNNs. We then present a comparison of the key numerical performance metrics of these works. We conclude by describing future research areas and applications of DNNs in massive MIMO receivers.
\end{abstract}
\begin{IEEEkeywords}
Data detection, deep learning, deep neural network, massive MIMO. 
\end{IEEEkeywords}

% \newpage 
\section{Introduction}
%\textit{Massive MIMO is a scaled up version of conventional MIMO system \cite{Rusek-SPMag-2013,LuLu-JSAC-2014,Zappone-TCOM2019}.}

As an integrated part of modern $5$G and emerging $6$G systems, massive MIMO offers several orders of magnitude enhancements in throughput and energy efficiency over conventional MIMO in existing $4$G systems \cite{LuLu-JSAC-2014,Rusek-SPMag-2013}. Through the use of large antenna arrays with tens to thousands of elements, massive MIMO enables the design of extremely narrow spatial beams that boost the desired signal power, resulting in considerable performance gains in terms of user coverage and system throughput. However, the increase in the dimension of massive MIMO and the corresponding increase in the number of served users adversely impact the complexity in its signal processing pipeline. For example, optimal maximum likelihood (ML) detection comes with a complexity that is exponential in the number of users. Low-complexity and near-optimal detection is thus crucial to fully realize the potential of massive MIMO system performance targets. 

For massive MIMO systems in which a base station equipped with a large array of antennas serving a large number of users simultaneously, low-complexity detectors such as zero-forcing (ZF) and linear minimum mean-squared error (LMMSE) may incur large performance gaps compared with the optimal ML detector. In contrast, near-optimal detection schemes, such as sphere decoding (SD), $K$--best SD (KSD), and fixed-complexity SD (FSD), may come at the cost of excessively high complexity \cite{nguyen2021application}. The algorithm deficits of these conventional approaches prompt the interesting prospect of applying deep learning (DL) for massive MIMO detection \cite{Zappone-TCOM2019}, in which the computational complexity is shifted to an offline training phase, enabling faster run time in the online detection phase.

The application of DL in communications has recently gained much attention. Several model-based deep neural network (DNN) architectures have been proposed for massive MIMO detection. The pioneering detection network (DetNet), introduced by Samuel \textit{et al.} \cite{Samuel-TSP2019}, has showcased the power of DL for MIMO data detection. A fast-convergence sparsely-connected neural network (FS-Net) has been recently proposed in \cite{Nhan-Lee-TWC-2020} as a simplified but optimized variant of DetNet. DetNet and FS-Net were both developed to mimic and optimize the iterative gradient descent algorithm. Another notable approach in DNN-based detection is based on the orthogonal approximate message passing (OAMP) algorithm \cite{OAMP-2017}, offering better performance as well as lower computational complexity compared to gradient descent-based algorithms. In particular, He \textit{et al.} introduced OAMP-Net2 \cite{He-TSP-2020} for data detection in both independent and identically distributed (i.i.d.) Gaussian and small-size correlated channels. Khani \textit{et al.} \cite{Khani-MIMO-NN-2019} proposed MMNet targeting data detection in correlated MIMO channels, and showed that it significantly outperforms OAMP-Net2. We note that all these detection networks are based on the deep unfolding technique \cite{Hershey-Unfolding-2014} and designed to optimize the free parameters of the underlying detection algorithms.

In this paper, we present a holistic framework for leveraging DNNs in massive MIMO data detection. We review the conventional MIMO detection algorithms and show how to incorporate the domain knowledge of these established algorithms into the development of DNN detectors, including DetNet, FSNet, OAMP-Net2, and MMNet.  We then present numerical results comparing key performance metrics of these works, including symbol error rate (SER) and run time. We conclude by describing future research areas and applications of DNNs in massive MIMO communications.  

\section{Background}
\subsection{Signal Model and MIMO Detection Problem}
We consider an uplink massive MIMO system, where the base station (BS) equipped with $N$ antennas serves $K$ single-antenna users. Note that the detectors presented in this article are also  applicable to multi-antenna users. The propagation channel from the users to the BS is modeled by a matrix $\mb{H}$ in which each entry represents the channel between a user and a receive antenna. We denote by $\mb{x}$ the vector of $K$ transmitted symbols associated with $K$ users, and we assume that these symbols are drawn from a discrete alphabet ${\mc{A}}$. The input-output relationship of the considered system is modeled as
\begin{eqnarray}\label{system-model}
    \mb{y} = \mb{H}\mb{x} + \mb{n},
\end{eqnarray}
where $\mb{y}$ is a vector of the received signals at the $N$ antennas of the BS, and $\mb{n}$ is a noise vector. Since the use of complex-valued parameters is uncommon in machine learning, we assume that all quantities in \eqref{system-model} are real-valued. This is also a matter of notational convenience, since a length-$n$ complex-valued vector is isomorphic to a length-$2n$ real-valued vector. In addition, a square complex-valued constellation of size $n^2$ (i.e., quadrature phase-shift keying (QPSK) and 16-quadrature amplitude modulation (16-QAM)) can be effectively represented by two independent real-valued alphabets of size $n$. The above model assumes a flat-fading or narrowband channel and the channel matrix is assumed to be known at the receiver. Our discussion can easily be extended to wideband channels using orthogonal frequency division multiplexing (OFDM). 

The task of MIMO detection is to determine the transmitted symbol vector $\mb{x}$ based on the received vector $\mb{y}$. The detection error is minimized by classifying the most likely $\mb{x}$ with the ML criterion when no {\it a priori} information is available. That is equivalent to finding the solution to the optimization problem $\min_{\mb{x} \in \mc{A}^K} \left\|\mb{y} -\mb{H} \mb{x} \right\|^2$. MIMO ML data detection is a combinatorial problem, and its complexity grows exponentially with the number of users $K$. Performing \emph{joint} detection of the entire symbol vector $\mb{x}$ is computationally expensive even for a small-scale MIMO system, and even more so for massive MIMO systems. % due to the enormous search space.
For example, the search space $\mc{A}^K$ grows to a set of $2^{32}$ candidates for a relatively \emph{modest} large-scale MIMO system supporting $8$ users with $16$-QAM. Thus, there is a need for near-optimal and reduced-complexity data detection algorithms that scale well to massive MIMO systems. To this end, we first review two typical classes of massive MIMO data detection schemes, namely, linear and nonlinear detectors with a comprehensive review in \cite{Albreem-ST-2019}.
%Interested readers are referred to a comprehensive review of these detection algorithms in \cite{Albreem-ST-2019}.

\subsection{Conventional MIMO/Massive MIMO  Data Detectors}
\subsubsection{Linear Data Detectors} 
Linear data detectors with low complexity are practical candidates for massive MIMO systems \cite{Rusek-SPMag-2013}. These detection schemes detect one symbol at a time while treating all the other symbols transmitted from the other users as interference. The estimated symbol is obtained from a linear combination of the received signals, which is then projected into the nearest symbol in the alphabet $\mathcal{A}$. The simplest of these is the matched-filter (MF) detector which aims at maximizing the energy of the signal of interest. A ZF detector targets elimination of the inter-user interference. Both schemes require relatively few computations, but they suffer from significant performance degradation due to the interference and/or noise enhancement. Unlike these two, an LMMSE detector tries to balance the enhancements in the signal of interest and the interference/noise. The LMMSE detector achieves the best performance among the three detectors, but it requires a matrix inversion, which can quickly result in excessive complexity for large-scale MIMO systems. 

While relatively simple to implement except the possible need for a matrix inversion, linear detectors can achieve good performance when the number of receive antennas is large enough compared to the number of users and the channel vectors from different users are independent \cite{Rusek-SPMag-2013,LuLu-JSAC-2014}. However, their performance deteriorates quickly when the number of users approaches the number of receive antennas or when the channel is ill-conditioned \cite{Albreem-ST-2019}, prompting the need for more sophisticated nonlinear detectors. 

\subsubsection{Nonlinear Detectors} 
SD is one of the most well-known nonlinear algorithms for MIMO detection. Similar to ML detection, SD attempts to find the optimal lattice point closest to $\mb{y}$. However, its search is limited to the points inside a hypersphere which is a subset of the feasible set $\setA^{K}$ and determined by a given radius. Each time a point lying inside the hypersphere is found, the search is further restricted by shrinking the sphere. When there is only one point in the sphere, the point becomes the final solution. The better optimized the sphere radius is, the better performance and/or complexity reduction can be achieved by SD \cite{nguyen2021application}. 

Approximate message passing (AMP) is a relatively low-complexity iterative signal recovery algorithm for large-scale linear systems. A variant of AMP, referred to as OAMP \cite{OAMP-2017}, has been exploited for MIMO data detection in recent papers. In OAMP, the recovered signal is updated via a nonlinear transformation of the previous iterate, which includes a linear estimator and a nonlinear denoiser. OAMP can attain near-optimal performance in few iterations. Except for a highly complicated matrix inversion in the linear estimator, OAMP can be a promising technique for massive MIMO detection. 
%In addition, it is computationally efficient, and is thus more promising for massive MIMO detection. 

%Compared to the belief propagation requiring $\mathcal{O}(KN)$ update messages per iteration, the AMP scheme with only $\mathcal{O}(K+N)$ messages has reduced complexity and is proven to perform asymptotically optimal for large-sized i.i.d.\ Gaussian channels. The OAMP is proposed for general channels; however, it requires performing matrix inversion in each iteration, causing significantly increased complexity compared to its conventional AMP counterpart.

% \footnote{CHRONOLOGY OF DETECTION TECHNIQUES FOR MASSIVE MIMO is presented in \cite{Albreem-ST-2019}.

% \begin{itemize}
% 	\item Matched filter (MF) detector:
% 	\item Zero-forcing (ZF) detector:
% 	\item Linear minimum mean-squared error (MMSE) detector:
% 	\item Sphere decoder: 
% 	\item Decision feedback equalization (DFE)
% 	\item Approximate message passing (AMP)
% 	\item Semidefinite relaxation (SDR)
% \end{itemize}}

The conventional MIMO detectors discussed above, especially those originally proposed for conventional small-sized MIMO systems, lead to a challenging performance-complexity tradeoff. Specifically, nonlinear detectors with near-optimal performance but high complexity may not be feasible for deployment in large-scale systems. On the other hand, the linear detectors with low complexity perform relatively poorly in large-scale systems where the numbers of users and receive antennas are comparable. This concern motivates recent research on DL for massive MIMO detection.

% Interestingly, this challenge can be efficiently overcome by leveraging the power of DL combined with the domain knowledge of existing MIMO detection schemes, as will be revealed in the following sections.

%\subsection{Classical Massive MIMO Detection}
% \cite{Albreem-ST-2019}

\section{DNN Detector for Linear MIMO Systems}
% \emph{One or two lines about machine learning: Machine learning uses algorithms to build analytical models, assisting computers learn from data. In last decade, machine learning has success stories in image classification, text recognition, speech recognition, natural language processing, robotics and so on. There are interesting applications of machine learning in communication system such as the detector design for the MIMO transmission}

In this section, we provide an overview of the design of DNN detectors in MIMO and massive MIMO systems. We first focus on the fundamentals of a DNN detector. We then review and analyze recent developments of DNN detectors in the literature.

\subsection{Fundamentals of DNN-based Data Detection}
A DNN model can be trained efficiently to provide reliable prediction/approximation of the transmitted signal vectors. It accepts the received signals $\mb{y}$ and channel information $\mb{H}$ as inputs and outputs an estimate $\hat{\mb{x}}$ of the transmitted signal vector $\mb{x}$. In this respect, $\hat{\mb{x}}$ can be modeled as the target of a nonlinear mapping $f(\mb{H},\mb{y};\bs{\theta})$, where $\bs{\theta}$ consists of parameters pertaining to the neural network. The fidelity of the mapping $f(\cdot)$, also known as the \emph{inference rule}, is measured by a cost function, which is defined as the mean squared error (MSE) between the estimate $\hat{\mb{x}}$ and the true transmitted signal $\mb{x}$.
The goal of a DNN detector is to design $f(\cdot)$ via the optimization of the  parameters $\bs{\theta}$ to minimize this cost function. 
% \begin{eqnarray}
%     L(\bs{\theta}) = \mathbb{E}_{\mb{H},\mb{x},\mb{y}\sim p_{\mr{data}}}\big[\|\mb{x}-\hat{\mb{x}}(\mb{H},\mb{y};\bs{\theta})\|^2\big].
% \end{eqnarray}
The data for training a DNN detector can be generated from the system model \eqref{system-model} with known prior distributions on the channel, the transmitted symbols, and the noise.

%Via the training of $f(\cdot)$ with learnable parameters $\bs{\theta}$, 
Most of the computational complexity of a DNN detector lies in the offline training phase. On the other hand, a DNN detector enables data detection with a much lower computational complexity at run time. This can be accomplished by performing the task in batch, offering polynomial time complexity in data detection based on simple matrix additions and multiplications. These operations are far simpler than the computationally expensive matrix inversions/pseudo-inversions or searching mechanisms that are performed in conventional linear or nonlinear detection algorithms. Furthermore, the DNN architectures and their batch operations are more natural for hardware implementation than hand-engineered algorithms, which is a critical distinction between the two.

An efficient DNN detector requires good designs across various aspects, including but not limited to the network architecture, input structure, and training strategy. In \cite{Samuel-TSP2019}, it was shown that a generic fully-connected DNN with only the received signals and channel coefficients as inputs leads to poor detection performance. In contrast, DetNet \cite{Samuel-TSP2019}, FS-Net \cite{Nhan-Lee-TWC-2020}, OAMP-Net2 and its predecessor OAMP-Net \cite{He-TSP-2020}, and MMNet \cite{Khani-MIMO-NN-2019} can achieve excellent performance in MIMO detection by exploiting not only the learning ability of DL but also the domain knowledge from hand-engineered data detection algorithms. All of these detectors follow an unfolding network architecture \cite{Hershey-Unfolding-2014}, allowing data detection to be performed in a layer-by-layer manner. The ingenuity of these architectures lies in the design of each layer, derived from well-developed data detection algorithms, leading to their differing performance and complexity. % in these DNN detectors.% Their designs and efficiencies will be discussed in detail as next.

\subsection{Gradient Descent-Based DNN Detectors}
\begin{figure*}[t]
	\small
\centering
\begin{tikzpicture}
[dot/.style={rectangle,draw=black,fill=white,inner sep=5pt,minimum size=.4cm,rounded corners=0pt}]
\draw [thick, rounded corners, dashed] (11.8,6.0) rectangle (22.9,10.6);

\node at (12.6,8.9) {$\mb{H}^T\mb{H}$};
%\draw [semithick,->] (12.2,8.9) to (12.7,8.9);
%\draw [semithick] (12.9,8.9) circle [radius=0.2];
%\node at (12.9,8.9) {$\times$};
\draw [semithick,->] (13.8,9.5) to (13.8,9.1);
\node at (13.9,9.7) {$\theta_{2\ell}$};
\draw [semithick,->] (13.1,8.9) to (13.6,8.9);
\draw [semithick] (13.8,8.9) circle [radius=0.2];
\node at (13.8,8.9) {$\times$};

\node at (12.5,7.3) {$-\mb{H}^T\mb{y}$};
%\draw [semithick,->] (12.2,7.3) to (12.7,7.3);
%\draw [semithick] (12.9,7.3) circle [radius=0.2];
%\node at (12.9,7.3) {$\times$};
\draw [semithick,->] (13.8,6.7) to (13.8,7.1);
\node at (13.9,6.45) {$\theta_{1\ell}$};
\draw [semithick,->] (13.1,7.3) to (13.6,7.3);
\draw [semithick] (13.8,7.3) circle [radius=0.2];
\node at (13.8,7.3) {$\times$};

\node at (12.7,8.1) {$\hat{\mb{x}}_{\ell}$};
\draw [semithick,->] (13.1,8.1) to (14.3,8.1);

%% r block
\draw [semithick,->] (13.8,8.1) to (13.8,8.7);
\draw [semithick,-] (14,8.9) to (14.5,8.9);
\draw [semithick,-] (14,7.3) to (14.5,7.3);
\draw [semithick] (14.5,8.1) circle [radius=0.2];
\node at (14.5,8.1) {$+$};
\node at (14.95,8.4) {$\mb{r}_{\ell}$};
\draw [semithick,->] (14.5,8.9) to (14.5,8.3);
\draw [semithick,->] (14.5,7.3) to (14.5,7.9);
\draw [semithick,->] (14.7,8.1) to (15.2,8.1);
\draw [semithick,->] (14.7,8.1) to (15.2,8.1);
\draw [thick] (15.2,7.8) rectangle (15.8,10.4);

\node at (12.7,10.1) {$\mb{v}_{\ell}$};
\draw [semithick,->] (13.1,10.1) to (15.2,10.1);
\node at (15.5,9.2) [rotate=90] {concatenate};

\draw [semithick,->] (15.8,9.1) to (16.3,9.1);
\draw [semithick] (16.5,9.1) circle [radius=0.2];
\node at (16.5,9.1) {$\times$};
\node at (16.6,8.25) {${\mb{W}}_{1\ell}$};
\draw [semithick,->] (16.5,8.5) to (16.5,8.9);

\draw [semithick,->] (16.7,9.1) to (17.2,9.1);
\draw [semithick] (17.4,9.1) circle [radius=0.2];
\node at (17.4,9.1) {$+$};
\node at (17.5,8.25) {${\mb{b}}_{1\ell}$};
\draw [semithick,->] (17.4,8.5) to (17.4,8.9);

\draw [semithick,->] (17.6,9.1) to (18,9.1);
\node[dot,draw=black,thick] at (18.3,9.1)  [text width=.22cm, align=center]{$\varrho$};
\draw [semithick,-] (18.6,9.1) to (19,9.1);
\draw [semithick,-] (19,8.1) to (19,10.1);

\draw [semithick,->] (19,10.1) to (19.4,10.1);
\draw [semithick] (19.6,10.1) circle [radius=0.2];
\node at (19.6,10.1) {$\times$};
\node at (19.7,9.25) {${\mb{W}}_{3\ell}$};
\draw [semithick,->] (19.6,9.5) to (19.6,9.9);
\draw [semithick,->] (19.8,10.1) to (20.3,10.1);
\draw [semithick] (20.5,10.1) circle [radius=0.2];
\node at (20.5,10.1) {$+$};
\node at (20.6,9.25) {${\mb{b}}_{3\ell}$};
\draw [semithick,->] (20.5,9.5) to (20.5,9.9);
\draw [semithick,->] (20.7,10.1) to (22,10.1);
\node at (22.4,10.05) {${\mb{v}}_{\ell+1}$};

\draw [semithick,->] (19,8.1) to (19.4,8.1);
\draw [semithick] (19.6,8.1) circle [radius=0.2];
\node at (19.6,8.1) {$\times$};
\node at (19.7,7.25) {${\mb{W}}_{2\ell}$};
\draw [semithick,->] (19.6,7.5) to (19.6,7.9);
\draw [semithick,->] (19.8,8.1) to (20.3,8.1);
\draw [semithick] (20.5,8.1) circle [radius=0.2];
\node at (20.5,8.1) {$+$};
\node at (20.6,7.25) {${\mb{b}}_{2\ell}$};
\draw [semithick,->] (20.5,7.5) to (20.5,7.9);
\draw [semithick,->] (20.7,8.1) to (20.99,8.1);
\node[dot,draw=black,thick] at (21.3,8.1)  [text width=.25cm, align=center]{$\psi$};
\draw [semithick,->] (21.6,8.1) to (22,8.1);
\node at (22.4,8.1) {$\hat{\mb{x}}_{\ell+1}$};

% \node at (17.5,6.5) {a. DetNet with trainable parameters};
% \node at (17.7,6.05) {};
%\node at (15.35,8.95) {$\hat{x}^{(\ell)}_{2}$};
\end{tikzpicture}
\caption{The $\ell$th layer of DetNet with trainable parameters $\bs{\theta}_{\ell} = \big\{\mb{W}_{1\ell},\mb{b}_{1\ell},\mb{W}_{2\ell},\mb{b}_{2\ell},\mb{W}_{3\ell},\mb{b}_{3\ell},\theta_{1\ell},\theta_{2\ell}\big\}$, a rectified linear unit (ReLU) $\varrho$, and a soft quantizer $\psi$.} 
\label{fig_DetNet}
\end{figure*}
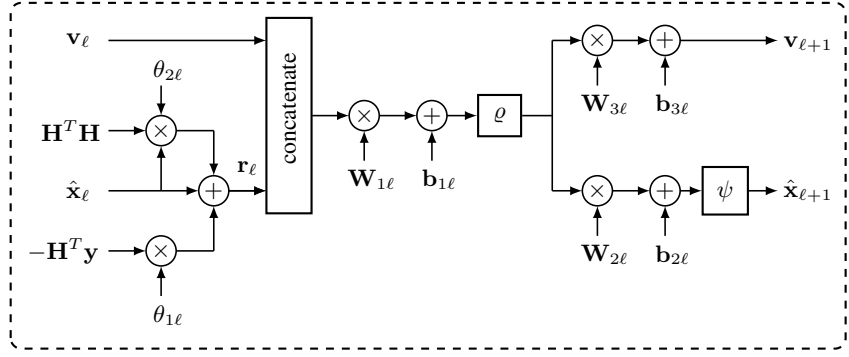

% \begin{figure}
% 	\centering
% 	\includegraphics[width=100mm]{soft_sign.eps}
% 	\caption{Soft quantizer function $\psi(x)$ for approximating the hard quantizer $\mc{P}_{\mc{A}}(x)$, where $\mc{A} = \{\pm 1\}$ for BPSK (QPSK) or $\mc{A} = \left\{\pm \frac{1}{\sqrt{5}},\pm \frac{3}{\sqrt{5}}\right\}$ for a unit energy $4$-ASK ($16$-QAM) alphabet. }
% \end{figure}

%The key role of knowledge domain in a successful design of DNN-based detection can be clearly shown in gradient descent-based detection schemes, which 
A gradient descent-based DNN detector incorporates the projected gradient descent (PGD) algorithm into the unfolding network architecture in an ingenious way. The network mimics the update process of the PGD algorithm and generates an estimated symbol vector at each layer. The operation at layer-$\ell$ is modeled by a nonlinear transformation $\hat{\mb{x}}_{\ell}= f_{\mathrm{gd}} \left( \hat{\mb{x}}_{\ell - 1} - \delta_{\ell} \mb{H}^T \mb{y} + \delta_{\ell} \mb{H}^T \mb{H} \hat{\mb{x}}_{\ell-1} \right)$, which accepts the output of the previous layer %(i.e., $\hat{\mb{x}}_{\ell - 1}$, $\mb{H}^T \mb{y}$, and $\mb{H}^T \mb{H}$)
and information from the channels and the received signals
as inputs. The network is trained to optimize the nonlinear transformation $f_{\mathrm{gd}}(\cdot)$ and the step sizes $\delta_{\ell}$, motivating the developments of DetNet and FS-Net.

\subsubsection{DetNet}

To learn the nonlinear projection $f_{\mathrm{gd}}(\cdot)$ of the original PGD method, DetNet employs a trainable parameter set %$\bs{\theta}$ %$=\big\{\mb{W}_{1\ell},\mb{b}_{1\ell},\mb{W}_{2\ell},\mb{b}_{2\ell},\mb{W}_{3\ell},\mb{b}_{3\ell},\theta_{1\ell},\theta_{2\ell}\big\}$, 
including the weights, biases, and step sizes. We illustrate the operation of the $\ell$th layer of DetNet in Fig. \ref{fig_DetNet}. In the DetNet architecture, a soft quantizer $\psi$ is introduced at the end of the layer to perform a soft element-wise quantization of the output $\hat{\mb{x}}_{\ell + 1}$. This ensures that the elements of $\hat{\mb{x}}_{\ell + 1}$ are in an appropriate range specified by the modulation scheme. In DetNet, the initial solution $\hat{\mb{x}}_{0}$ is set to all-zero vector, which is then updated over $L$ layers of the DNN to approach the true transmit signal vector by minimizing the loss function $\sum_{\ell=1}^{L} \log(\ell) \norm{\mb{x} - \hat{\mb{x}}_{\ell}}^2$. The final solution of DetNet is obtained by a hard quantization of the last layer's outputs (i.e., $\hat{\mb{x}}_{L}$) to the nearest symbols in the alphabet $\mathcal{A}$.

The network architecture and operations of DetNet exhibit the following potential issues:
\begin{itemize}
    \item The value of the  loss function of one layer in DetNet is added to the total loss of the network with a discounted weight, while the solution predicted in one layer is obtained using only the connections in that layer and the input passed from the previous layer. Sophisticated features cannot be extracted within one layer, implying that the loss function of DetNet limits the learning ability of multiple hidden layers in a general DNN. Furthermore, it is evident that this loss function only minimizes the total loss of all the layers. However, it does not minimize the number of required layers to accelerate the training and prediction \cite{Nhan-Lee-TWC-2020}.
    
    \item As seen in Fig. \ref{fig_DetNet}, an intermediate signal vector $\mb{v}_{\ell}$ is concatenated with $\hat{\mb{x}}_{\ell} - \delta_{\ell  +1} \mb{H}^T \mb{y} + \delta_{\ell +1} \mb{H}^T \mb{H} \hat{\mb{x}}_{\ell}$ to form the inputs that are processed by the network connections. Although $\mb{v}_{\ell}$ helps to overcome the limitations of the loss function \cite{Samuel-TSP2019}, it obviously enlarges the size of the input vector and additionally requires a sub-network associated with the trainable parameter set $\{\mb{W}_{3\ell},\mb{b}_{3\ell}\}$. This makes DetNet computationally expensive. Moreover, the use of different step sizes (i.e., $\theta_{1\ell},\theta_{2\ell}$) in DetNet is not clearly motivated or suggested by the PGD procedure in $f_{\mathrm{gd}}$. We note that both $\mb{v}_{\ell}$ and $\theta_{2\ell}$ can be removed in refined versions of DetNet. %, as will be detailed latter.
\end{itemize}

Despite the above limitations, DetNet offers several performance advantages. In simulations for a massive MIMO system with $60$ receive and $30$ transmit antennas and binary phase-shift keying (BPSK), DetNet exhibits a $2$-dB performance gain over the ZF detector and performs very close to the SD scheme with much lower complexity. This justifies the potential of DNNs in estimating the symbols transmitted via fading channels and observed by a noisy receiver. %  DetNet can be considered as one of the most important finding the literature attempting to optimize the performance-complexity tradeoff in MIMO and massive MIMO data detection, creating rooms for further improvements and applications.

%. Furthermore, although the performance of DetNet is shown to be good for the case $N \gg K$, subsequent work \cite{Nhan-Lee-TWC-2020} showed the network performance to be far from optimal for square systems, i.e., $N \approx K$.  

\begin{figure}
    \centering
    %% FS-Net
    \begin{tikzpicture}
    [dot/.style={rectangle,draw=black,fill=white,inner sep=5pt,minimum size=.4cm,rounded corners=0pt}]
    \draw [thick, rounded corners, dashed] (10.8,7.6) rectangle (18.6,10.8);
    \node at (11.8,9.7) {$\hat{\mb{x}}_{\ell}$};
    \draw [semithick,->] (12.2,9.7) to (14.5,9.7);
    \draw [semithick,->] (12.9,9.7) to (12.9,9.1);
    
    \node at (11.7,8.9) {$\mb{H}^T\mb{H}$};
    \draw [semithick,->] (12.2,8.9) to (12.7,8.9);
    \draw [semithick] (12.9,8.9) circle [radius=0.2];
    \node at (12.9,8.9) {$\times$};
    \draw [semithick,->] (13.1,8.9) to (13.6,8.9);
    \draw [semithick] (13.8,8.9) circle [radius=0.2];
    \node at (13.8,8.9) {$+$};
    \node at (11.6,8.1) {$-\mb{H}^T\mb{y}$};
    \draw [semithick,-] (12.2,8.1) to (13.8,8.1);
    \draw [semithick,->] (13.8,8.1) to (13.8,8.7);
    
    \draw [semithick,->] (14,8.9) to (14.5,8.9);
    \draw [semithick] (14.7,8.9) circle [radius=0.2];
    \node at (14.7,8.9) {$\times$};
    \draw [semithick,->] (14.7,8.3) to (14.7,8.7);
    \node at (14.7,8.0) {${\mb{W}}_{1\ell}$};
    
    \draw [semithick] (14.7,9.7) circle [radius=0.2];
    \node at (14.7,9.7) {$\times$};
    \draw [semithick,->] (14.7,10.3) to (14.7,9.9);
    \node at (14.7,10.5) {${\mb{W}}_{2\ell}$};
    
    \draw [semithick,->] (14.9,9.7) to (15.3,9.7);
    \draw [semithick,-] (14.9,8.9) to (15.5,8.9);
    \draw [semithick,->] (15.5,8.9) to (15.5,9.5);
    \draw [semithick] (15.5,9.7) circle [radius=0.2];
    \node at (15.5,9.7) {$+$};
    \node at (15.5,10.5) {${\mb{b}}_{\ell}$};
    \draw [semithick,->] (15.5,10.3) to (15.5,9.9);
    
    \draw [semithick,->] (15.7,9.7) to (16.25,9.7);
    \node at (15.95,10) {$\mb{r}_{\ell}$};
    \node[dot,draw=black,thick] at (16.55,9.7)  [text width=.3cm, align=center]{$\psi$};
    \draw [semithick,->] (16.9,9.7) to (17.3,9.7);
    \node at (17.7,9.7) {$\hat{\mb{x}}_{\ell+1}$};
    \end{tikzpicture}
    \caption{The $\ell$th layer of FS-Net with trainable parameters $\bs{\theta}_{\ell} = \big\{\mb{W}_{1\ell},\mb{W}_{2\ell},\mb{b}_{\ell}\big\}_{\ell=1}^L$ with diagonal $\mb{W}_{1\ell}$ and $\mb{W}_{2\ell}$ and a soft quantizer $\psi$.} 
    \label{fig_FSNet}
\end{figure}
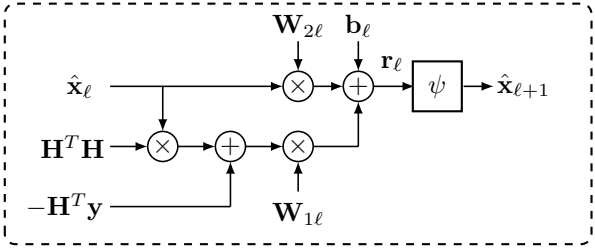

\subsubsection{FS-Net}
% Several network architecture have been proposed based on the DetNet. For example, a DNN detector called Twin-DNN uses two parallel DetNets, one accepts the ZF solution (reviewed earlier) as the input, while the other has a random initial solution. The solution of Twin-DNN is the more accurate of the two output vectors of the two DetNets, motivated by ensemble learning. As a result, Twin-DNN improves the performance at the cost of approximately double the complexity of DetNet. Another variant of DetNet is Cascade-Net, which was proposed for single-antenna systems. In Cascade-Net, a DNN is cascaded with a ZF preprocessor to prevent the network from converging to a saddle point or local minimum point to achieve better performance than DetNet and the classical ZF detector.

FS-Net is proposed in \cite{Nhan-Lee-TWC-2020} to overcome the limitations of DetNet. It achieves not only a considerable complexity reduction thanks to a simple network architecture (as seen in Fig. \ref{fig_FSNet}) but also significant performance improvement compared to DetNet \cite{Nhan-Lee-TWC-2020}. These gains are obtained thanks to the following improvements:
\begin{itemize}
    \item FS-Net does not require the intermediate vector $\mb{v}_{\ell}$ and only uses one trainable step size. Clearly, this simplifies the network structure and reduces the number of trainable parameters approximately by half, facilitating a better training while also being more computationally efficient than the original DetNet.
    
    \item In FS-Net, pair-wise connections between the input and output nodes are deployed instead of full connections as in DetNet. This is motivated by the fact that in $f_{\mathrm{gd}}$, an element of the output $\hat{\mb{x}}_{\ell+1}$ only depends on the corresponding element of $\hat{\mb{x}}_{\ell}$. The pair-wise connections significantly reduce the number of trainable parameters.

    \item Finally, FS-Net employs an optimized loss function to accelerate the convergence in the training phase. The new loss function takes into account the correlation between the output of each layer and the label (i.e., the true transmitted signal vectors), thus ensuring that $\hat{\mb{x}}_{\ell}$ can reach $\mb{x}$ with fewer layers, compared to DetNet.
\end{itemize}

% Numerical comparisons between the performance of  FS-Net and DetNet will be provided latter in our simulation results.

\begin{figure}
	\small
\centering
%% OAMPnet
\begin{tikzpicture}
[dot/.style={rectangle,draw=black,fill=white,inner sep=5pt,minimum size=.4cm,rounded corners=0pt}]
\draw [thick, rounded corners, dashed] (10.5,2.6) rectangle (19.3,10.6);
\node at (11.8,10) {$\hat{\mb{x}}_{\ell}$};
\draw [semithick,->] (12.2,10) to (14.8,10);
\draw [semithick,->] (12.9,10) to (12.9,9.1);

\node at (11.7,8.9) {$-\mb{H}$};
\draw [semithick,->] (12.2,8.9) to (12.7,8.9);
\draw [semithick] (12.9,8.9) circle [radius=0.2];
\node at (12.9,8.9) {$\times$};
\draw [semithick,->] (13.1,8.9) to (13.6,8.9);
\draw [semithick] (13.8,8.9) circle [radius=0.2];
\node at (13.8,8.9) {$+$};
\node at (11.9,8.1) {$\mb{y}$};
\draw [semithick,-] (12.2,8.1) to (13.8,8.1);
\draw [semithick,->] (13.8,8.1) to (13.8,8.7);

\node[dot,draw=black,thick] at (15.05,8.9)  [text width=.7cm, align=center]{$\gamma_{\ell}\mb{W}_{\ell}$};
\draw [semithick,->] (14,8.9) to (14.5,8.9);
\draw [semithick,->] (15.6,8.9) to (16.08,8.9);
\draw [semithick,-] (14.2,8.9) to (14.2,8.1);
\draw [semithick,-]  (14.2,8.1) to (16.7,8.1);
\draw [semithick,->] (16.7,8.1) to (16.7,8.53);
\node[dot,draw=black,thick] at (16.7,8.9)  [text width=.9cm, align=center]{$\tau_{\ell}^2(\theta_{\ell})$};
\draw [semithick,->] (16.7,9.276) to (16.7,9.63);
\node at (15.85,8.4) {$v^2_{\ell}$};

%\draw [semithick,-] (14.9,8.9) to (15.5,8.9);
\draw [semithick,->] (15,9.25) to (15,9.8);
\draw [semithick] (15.,10) circle [radius=0.2];
\node at (15,10) {$+$};
\draw [semithick,->] (15.2,10) to (15.77,10);
\node at (15.45,10.3) {$\mb{r}_{\ell}$};
\node[dot,draw=black,thick] at (16.7,10)  [text width=1.5cm, align=center]{$\eta_{\ell}(\cdot;\phi_{\ell},\xi_{\ell})$};
\draw [semithick,->] (17.62,10) to (18.02,10);
\node at (18.46,10) {$\hat{\mb{x}}_{\ell+1}$};

\node at (14.9,7.5) {OAMP-Net2 with trainable parameters};
\node at (14.9,7.05) {$\bs{\theta} = \big\{\gamma_{\ell},\theta_{\ell},\phi_{\ell},\xi_{\ell}\big\}_{\ell=1}^L$ and an element-wise denoiser $\eta_{\ell}(\cdot)$}; 
% {with an element-wise denoiser $\eta_{\ell}(\cdot)$ depending on $\bs{\theta}_{2\ell}$};
% \end{tikzpicture}
% \\
% \vspace{0.3cm}
% %% MMnet
% \begin{tikzpicture}
% [dot/.style={rectangle,draw=black,fill=white,inner sep=5pt,minimum size=.4cm,rounded corners=0pt}]
% \draw [thick, rounded corners, dashed] (10.5,6.6) rectangle (19.3,10.3);
\node at (11.8,5.7) {$\hat{\mb{x}}_{\ell}$};
\draw [semithick,->] (12.2,5.7) to (15.3,5.7);
\draw [semithick,->] (12.9,5.7) to (12.9,5.1);

\node at (11.7,4.9) {$-\mb{H}$};
\draw [semithick,->] (12.2,4.9) to (12.7,4.9);
\draw [semithick] (12.9,4.9) circle [radius=0.2];
\node at (12.9,4.9) {$\times$};
\draw [semithick,->] (13.1,4.9) to (13.6,4.9);
\draw [semithick] (13.8,4.9) circle [radius=0.2];
\node at (13.8,4.9) {$+$};
\node at (11.9,4.1) {$\mb{y}$};
\draw [semithick,-] (12.2,4.1) to (13.8,4.1);
\draw [semithick,->] (13.8,4.1) to (13.8,4.7);

\draw [semithick,->] (14,4.9) to (14.5,4.9);
\draw [semithick] (14.7,4.9) circle [radius=0.2];
\node at (14.7,4.9) {$\times$};
\draw [semithick,->] (14.7,4.3) to (14.7,4.7);
\node at (14.8,4.0) {$\bs{\Theta}_{1\ell}$};

\draw [semithick,-] (14.9,4.9) to (15.5,4.9);
\draw [semithick,->] (15.5,4.9) to (15.5,5.5);
\draw [semithick] (15.5,5.7) circle [radius=0.2];
\node at (15.5,5.7) {$+$};
%\node at (15.25,9.35) {$-$};
\draw [semithick,->] (15.7,5.7) to (16.12,5.7);
\node at (15.85,6) {$\mb{r}_{\ell}$};
\node[dot,draw=black,thick] at (16.9,5.7)  [text width=1.2cm, align=center]{$\eta_{\ell}(\cdot;\bs{\theta}_{2\ell})$};
\draw [semithick,->] (17.68,5.7) to (18.08,5.7);
\node at (18.52,5.7) {$\hat{\mb{x}}_{\ell+1}$};

\node at (14.9,3.5) {MMNet with trainable parameters};
\node at (14.85,3.05) {$\bs{\theta} = \big\{\bs{\Theta}_{1\ell},\bs{\theta}_{2\ell}\big\}_{\ell=1}^L$ and an element-wise denoiser $\eta_{\ell}(\cdot)$};
\end{tikzpicture}
\caption{The $\ell$th layer of OAMP-Net2 and MMNet with trainable parameters $\bs{\theta}$.}
\label{OAMP-Net-figure}
\end{figure}
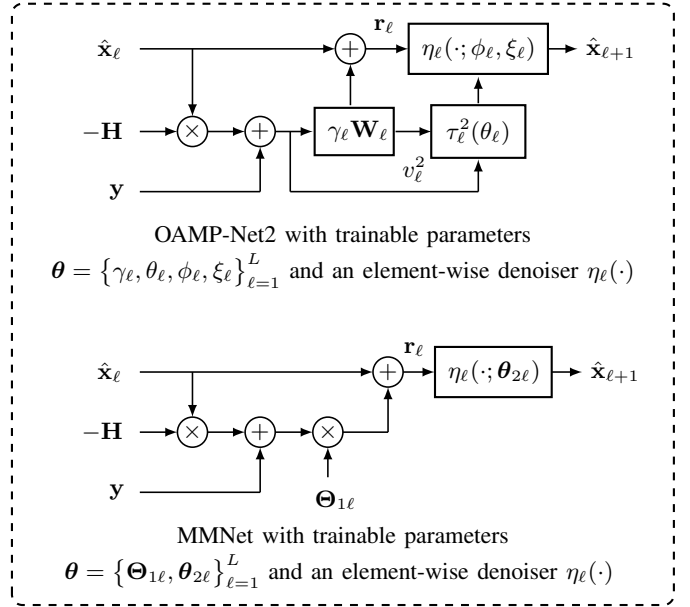

\subsection{Approximate Message Passing-based DNN Detectors}
% \cite{He-TSP-2020}
% \cite{Khani-MIMO-NN-2019}
In this section, we review another prominent group of DNN detectors, consisting of OAMP-Net2 \cite{He-TSP-2020} and MMNet \cite{Khani-MIMO-NN-2019}. Similar to their PGD-based counterparts, OAMP-Net2 and MMNet follow the unfolding technique \cite{Hershey-Unfolding-2014}. However, the major distinction between these two groups is the domain knowledge leveraged for constructing the layered architecture. The DNN detectors in this group are based on the iterative OAMP signal recovery algorithm. 

The OAMP framework sequentially invokes a linear estimator and a nonlinear denoiser to refine the recovered signal. At iteration $\ell$, it computes a linear estimate $\mb{r}_{\ell} = \hat{\mb{x}}_{\ell} + \mb{W}_{\ell}(\mb{y} - \mb{H}\hat{\mb{x}}_{\ell})$, using the estimated signal from the previous iteration and a linear estimator $\mb{W}_{\ell}$. The linear estimate is then passed through a nonlinear denoiser $\eta\big(\mb{x}|\mb{r}_{\ell},\tau_{\ell}\big)$ that provides a \emph{divergence-free} estimate $\hat{\mb{x}}_{\ell+1}$. This nonlinear denoiser is an affine function of the posterior mean $\tilde{\eta}\big(\mb{x}|\mb{r}_{\ell},\tau_{\ell}\big) = \mathbb{E}[\mb{x}|\mb{r}_{\ell} = \mb{x}+\tau_{\ell}\mb{z}]$, where $\tau_{\ell}^2$ is treated as the error variance and $\mb{z}$ is an i.i.d.\ standard Gaussian distributed error vector after the linear estimation. To improve the performance of OAMP as a data detection algorithm, OAMP-Net2 and MMNet were proposed to leverage the learning ability of DNNs for optimizing the free parameters in the linear estimator and the nonlinear denoiser.

\subsubsection{OAMP-Net2}
OAMP-Net2, as illustrated in Fig. \ref{OAMP-Net-figure}, and its predecessor OAMP-Net strictly follows the OAMP framework. Specifically, He \textit{et al.} \cite{He-TSP-2020} proposed the training of four variables $\{\gamma_{\ell}, \theta_{\ell}, \phi_{\ell}, \xi_{\ell} \}$ at each layer to form the linear estimate $\hat{\mb{x}}_{\ell} + \gamma_{\ell} \mb{W}_{\ell}(\mb{y} - \mb{H}\hat{\mb{x}}_{\ell})$ and the denoiser $\eta\big(\cdot;\phi_{\ell},\xi_{\ell},\tau_{\ell} \big) = \phi_{\ell} \tilde{\eta}\big(\mb{x}|\mb{r}_{\ell},\tau_{\ell} \big) - \xi_{\ell} \mb{r}_{\ell}(\gamma_{\ell})$. The trained parameters can significantly improve the accuracy and convergence of the nonlinear estimator. Specifically, $\{\gamma_{\ell}, \theta_{\ell} \}$ can improve the accuracy in estimating the prior mean $\mb{r}_{\ell}$ and variance $\tau_{\ell}^2(\theta_{\ell})$ in the nonlinear estimator. At the same time, $\phi_{\ell}$ and $\xi_{\ell}$ are trained to achieve a better divergence-free nonlinear estimator $\hat{\mb{x}}_{\ell+1}$ than the analytical solution in \cite{OAMP-2017}.
 
OAMP-Net2 achieves an impressive performance improvement compared to the conventional linear/nonlinear detectors. Specifically, a numerical example for an $8 \times 8$ MIMO system with i.i.d.\ Rayleigh fading channels shows that it can perform $5$-dB and $10$-dB better than the classical OAMP and LMMSE schemes \cite{He-TSP-2020}. However, like OAMP, OAMP-Net2 is strictly based on the assumption of unitarily-invariant channels. Therefore, it has a significant performance loss for realistic correlated channel models \cite{Khani-MIMO-NN-2019}. Furthermore, OAMP-Net2 imposes high complexity (even higher than the classical OAMP) due to the additional trainable parameters.

\subsubsection{MMNet}
%\textcolor{red}{May need a couple of sentences to mention MMNet-iid for iid channel and MMNet for specific channel realization.}
Khani \textit{et al.} proposed MMNet \cite{Khani-MIMO-NN-2019} to overcome the limitations of OAMP-Net. Similar to OAMP-Net and OAMP-Net2, MMNet unfolds the iterative update of the linear and nonlinear estimators. However, a significant improvement is made to overcome the poor performance of the OAMP algorithm for correlated channels. As illustrated in Fig. \ref{OAMP-Net-figure}, MMNet can be summarized as follows:
\begin{itemize}
    \item Matrix $\mb{W}_{\ell}$, which represents the linear transformation in the linear estimator of OAMP and OAMP-Net2, is cast as a trainable matrix variable $\boldsymbol{\Theta}_{1 \ell}$ in MMNet. This allows the linear estimator $\mb{r}_{\ell}$ to include more trainable parameters that can be optimized for each channel realization%., thereby facilitating online training with adaptation. 
    Moreover, it also avoids the matrix inversion in $\mb{W}_{\ell}$, required by OAMP and OAMP-Net2.
    \item To handle the cases in which different transmitted symbols have different noise levels, the error variance $\tau^2_{\ell}$ is parameterized by a length-$K$ vector $\boldsymbol{\theta}_{2\ell}$, corresponding to the estimated error variances for the $K$ users at the denoiser input.
\end{itemize}

%As a result, the update process in the $\ell$th layer of MMNet is given as $\mb{r}_{\ell} =  \hat{\mb{x}}_{\ell} + \boldsymbol{\Theta}_{1 \ell} (\mb{y} - \mb{H}\hat{\mb{x}}_{\ell})$ and $\hat{\mb{x}}_{\ell+1} = \eta\big(\mb{x}|\mb{r}_{\ell}(\boldsymbol{\Theta}_{1 \ell}),\tau_{\ell} (\boldsymbol{\theta}_{2\ell}) \big)$. It is observed that more flexibility is introduced to the linear estimator as well as the denoiser in MMNet, 

MMNet offers more flexibility in designing the linear estimator and the denoiser, compared to the OAMP algorithm. Simulation results in \cite{Khani-MIMO-NN-2019} showed that MMNet outperforms OAMP-Net by $3$-dB and reduces the computational complexity by a factor of $10$--$15$ for practical 3GPP channels. It is, however, noted that MMNet requires retraining for each channel realization. A simplified version of MMNet, called MMNet-iid, was also proposed in \cite{Khani-MIMO-NN-2019} for detection with i.i.d.\ Gaussian channels. In MMNet-iid, the trainable matrix/vector $\boldsymbol{\Theta}_{1 \ell}$ and $\boldsymbol{\theta}_{2\ell}$ are replaced by $\theta_{1\ell}\mb{H}^T$ and $\theta_{2\ell}$, respectively, where $\theta_{1\ell}$ and $\theta_{2\ell}$ are trainable scalars. 
% Note that for the case that the noise are the same for all the transmitted symbol, 

\section{Numerical Examples and Discussion}
\subsection{Numerical Examples}
\begin{figure}[t!]
	\centering
	\includegraphics[width=1\linewidth]{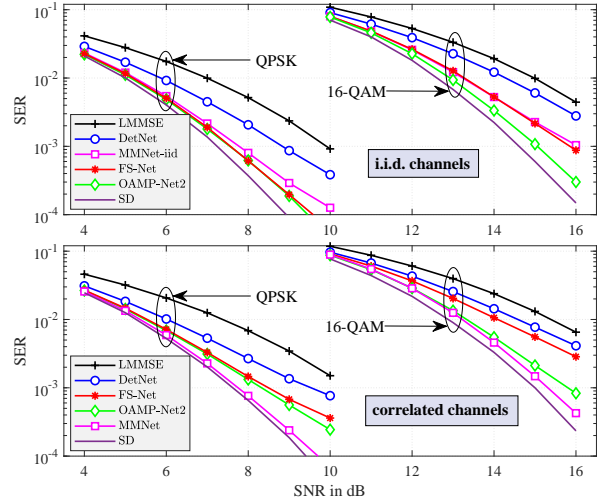}
	\caption{Performance comparison between LMMSE, SD, and different detection networks with i.i.d. and correlated channels.}
	\label{fig_SER}
\end{figure}
\begin{figure}[t!]
	\centering
	\includegraphics[width=1\linewidth]{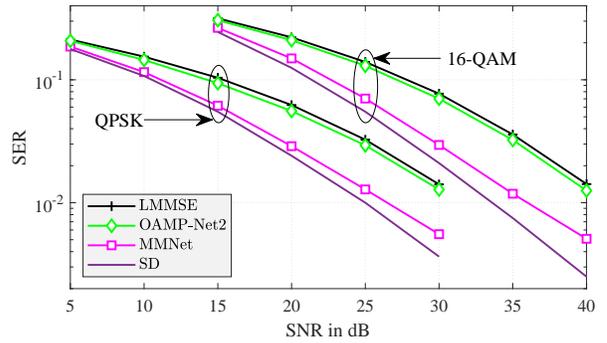}
	\caption{Performance comparison between LMMSE, SD, and different detection networks with realistic 3GPP channels.}
	\label{fig_SER_3gpp}
\end{figure}
\begin{table*}[t!]
\centering
\caption{Computational complexity comparison in terms of average run time (seconds). \label{table_runtime}}
\renewcommand{\arraystretch}{1.4}
\begin{tabular}{|l|c|c|c|c|c|c|c|c|c|}
\hline
\multirow{2}{*}{} & \multirow{2}{*}{\textbf{LMMSE}} & \multirow{2}{*}{\textbf{FS-Net}} & \multirow{2}{*}{\textbf{MMNet-iid}} & \multirow{2}{*}{\textbf{OAMP-Net2}} & \multirow{2}{*}{\textbf{DetNet}} & \multirow{2}{*}{\textbf{SD}} & \multicolumn{3}{c|}{\textbf{MMNet (include training time)}}\\ \cline{8-10} & & & & & & & \multicolumn{1}{l|}{$10$ epochs} & \multicolumn{1}{l|}{$100$ epochs} & \multicolumn{1}{l|}{$500$ epochs} \\ \hline
\textbf{QPSK} & $0.4\times10^{-6}$ & $2\times10^{-6}$ & $11\times10^{-6}$ & $13\times10^{-6}$ & $15\times10^{-6}$ & $>5\times10^{-4}$ & $1.4$ & $12$ & $60$ \\ \hline
\textbf{16-QAM}   & $0.4\times10^{-6}$ & $3\times10^{-6}$ & $23\times10^{-6}$ & $25\times10^{-6}$ & $50\times10^{-6}$ & $>6\times10^{-4}$ & $2.2$ & $19$ & $94$ \\ \hline
\end{tabular}
\end{table*}

Figs.~\ref{fig_SER} and~\ref{fig_SER_3gpp} provide performance comparisons between the discussed detection networks (i.e., DetNet, MMNet, FSNet, and OAMPNet2) and the conventional LMMSE and SD detectors. We consider $(K,N) = (16,32)$ and set $L=10$ for QPSK and $L=15$ for $16$-QAM. In the training phase, we set the learning rate to $10^{-3}$ and the batch training size to $1000$. Simulations were implemented on a standard Intel Xeon CPU E3-1270 v5, 3.60 GHz with $16$-GB RAM, using the Tensorflow library. It should be noted that except for MMNet, all the other detection networks are trained offline. MMNet was designed to be trained online (i.e., it has to be retrained whenever the channel matrix $\mathbf{H}$ changes).

%\textcolor{red}{Need more details here! Performance gaps! Break into two paragraphs one for i.i.d., one for correlated channel (mention which correlated channel modeled being used!} 

The performance comparison in the upper part of Fig.~\ref{fig_SER} is for the case of i.i.d.\ Rayleigh fading channels. It shows that the DNN-based detectors outperform the LMMSE scheme. Among the considered DNN-based detectors, DetNet provides the worst performance. Compared to LMMSE, the gain of DetNet is only about $1$-dB and $0.5$-dB for QPSK and $16$-QAM, respectively. The OAMP-Net2 detector provides the best performance (quite close to that of the SD method) with a $2$-dB gain compared to LMMSE in both the cases of QPSK and $16$-QAM. FS-Net performs as well as OAMP-Net2 for the case of QPSK, but worse for $16$-QAM. The performance of MMNet-iid is between DetNet and OAMP-Net2.

The lower part of Fig.~\ref{fig_SER} presents a performance comparison for the case of spatially correlated channels. We assume that the channels from different users to the BS are uncorrelated but the channels from a given user to the receive antennas are spatially correlated and follow a typical urban channel model as described in \cite{Pedersen-TVT-2000}. It is also observed that the DNN-based detectors outperform the LMMSE scheme. Among the considered DNN-based detectors, DetNet also provides the least performance gain at about $1$-dB and $0.5$-dB over LMMSE for the case of QPSK and $16$-QAM, respectively. MMNet achieves the lowest SER (also quite close to that of the SD method) in this correlated channel scenario thanks to its online training strategy, but with the cost of excessively high computational complexity. In contrast, the other DNN-based detectors are trained offline before the online detection (re-training is not required) and, thus they have lower computational complexities compared to MMNet. Note that the complexity of offline training is generally ignored in the literature \cite{nguyen2021application, Nhan-Lee-TWC-2020}. The gain of MMNet compared to LMMSE is significant (more than $2$-dB). While FS-Net and OAMP-Net2 give similar performance for QPSK, FS-Net performs worse than OAMP-Net2 for $16$-QAM, similar to what was observed in i.i.d.\ channels. 

Realistic 3GPP channels are considered in Fig.~\ref{fig_SER_3gpp}, where the QuaDRiGa 3GPP model~\cite{Jaeckel2014Quadriga} is adopted. 
We observed that the training process of DetNet and FS-Net did not converge with this channel model (a similar observation was reported in~\cite{Khani-MIMO-NN-2019}). Therefore, we compare the two detection networks OAMP-Net2 and MMNet with LMMSE and SD. 
MMNet performs closest to SD and much better than OAMP-Net2 and LMMSE. As explained earlier, this is due to the online training strategy of MMNet.

Table~\ref{table_runtime} compares the computational complexity of the detection methods in terms of average run time. It is obvious that LMMSE has the lowest complexity since it is a linear detector. The complexity of FS-Net is the lowest among the network detectors. The run times of MMNet-iid and OAMP-Net2 are longer than that of FS-Net, since they use more complex denoisers and OAMP-Net2 requires a matrix inversion in each layer. Among the DNN detectors that use offline training, DetNet has the longest run time because its layered structure is more sophisticated with many parameters and the input of each layer is also lifted to a much higher dimension. 
All the offline-training DNN detectors run faster than the SD detector. The computational complexity of MMNet is much higher than that of the other detectors, since it must be trained online.

\subsection{Discussion}
DNN-based detection requires an offline training phase. The resulting trained DNN model is saved at the base station for online application. Once the DNN is deployed, the base-station does not require further training data, except for MMNet which uses online retraining. The following is a summary of the advantages and disadvantages of the presented detection networks:
\begin{itemize}
    \item \textit{DetNet} is better than LMMSE but computationally expensive due to its sophisticated structure.
    \item \textit{FS-Net} has lower complexity than \textit{DetNet}, but its performance is degraded with large constellations. Both \textit{FS-Net} and \textit{DetNet} may not converge for certain practical channels.
    \item \textit{OAMP-Net2} performs well with i.i.d. and correlated channels, but not as well with realistic channels and is computationally expensive due to the use of matrix inversions.
    \item \textit{MMNet}, while working well with any channel model, is very computationally expensive due to the need for online retraining. The simplified version  \textit{MMNet-iid} with offline training performs well for its targeted i.i.d. channels.
\end{itemize}
Given the above discussion, it is clear that the development of more efficient, low-complexity, and universally-applicable DNN-based detectors is of significant interest.

\section{Open Research Problems}
\subsection{Learning to Learn the MIMO Detector}
The aforementioned DNN detectors tune their inference rules based on the training data. %However, none of these detectors take into account the knowledge of the underlying distribution of the training data. 
If there is a change in the data distribution (e.g., spatially correlated channel or sparse channel, a new mapping for the transmitted symbols, or a spatially correlated noise model), the trained DNN detector may become obsolete. Retraining the DNN detector from scratch for each new data distribution may not be feasible. This issue prompts the consideration of meta-learning in the DNN detector design.

Meta-learning, also known as ``learning to learn,'' aims to design a model that learns from the output of other learning models using previously observed tasks. A notable meta-learning approach is to train the meta-learner's initial parameters such that the model has maximal performance on new tasks with just a few gradient update steps \cite{Finn-MAML-2017}. In the context of DNN-based detection, it would be interesting to investigate how to apply meta-learning to pre-train the weights of the DNN detector to a good initialization point that generalizes well to new underlying data distributions.

\subsection{Channel Estimation and Channel Decoding}
A DNN detector requires knowledge of the channel, which must be estimated before the data detection phase. It would be interesting to investigate the performance of the DNN detectors with potential channel estimation mismatch. In addition, the novel model-based DNN architectures can be designed to carry out both channel estimation and data detection tasks. 

Channel encoding/decoding is another integral part of communications systems. A well performing code typically requires soft inputs from the demodulator. Thus, it is important for a DNN detector to provide soft detection outputs to the channel decoder. In addition, the DNN detector should be able to accept soft outputs from the channel decoder as prior information for the data symbol vector. This implementation would allow turbo-like joint MIMO detection and channel decoding with DNNs.

\subsection{DNN-based Detection for Nonlinear MIMO Channels}
The majority of the proposed DNN-based detectors in the literature tackle the detection problem in linear MIMO channels. However, a cost-efficient and energy-efficient massive MIMO system may use non-ideal hardware that is prone to impairments and nonlinear distortions. A DNN detector for massive MIMO systems with one-bit ADCs, proposed in a recent work \cite{Ly-TWC-2020}, has shown significant performance gain over algorithm-based approaches. For massive MIMO systems that exhibit nonlinear power amplifiers and phase noise, developing novel DNN detectors is an open research direction.

%\subsection{\textcolor{red}{Resource Allocation \& Physical Layer Security}}

\section{Conclusion}
We have reviewed several recent developments in DNN-based massive MIMO detection. By imitating the iterations in established MIMO detection algorithms with a predetermined number of layers, a DNN detector with learned and fine-tuned parameters can offer fewer detection errors with lower computational complexity at run time. We believe that DNN-based detection can contribute to the development of low-complexity technologies for modern and emerging wireless networks.

\end{document}